# Resummed Predictions for the Structure Function $F_2$ at Small $x$[*]

F. Hautmann

Cavendish Laboratory
Department of Physics, University of Cambridge
Cambridge CB3 0HE, UK

## Abstract

We report the results of including resummed splitting functions in the QCD evolution equations at small $x$, and discuss the predictions that follow for the deep inelastic structure functions.

---

[*]Talk given at the XXX Rencontres de Moriond, Les Arcs, March 1995.

# 1. Introduction

Present experiments at the HERA $ep$ collider and fixed target accelerators probe deep inelastic lepton-hadron scattering in the region of low values of the Bjorken variable $x$ [1-3]. As $x$ decreases, the scattering process approaches the high-energy regime, in which the perturbative QCD dynamics is dominated by multiple gluon exchange, and described in leading order by the BFKL equation [4]. The question then arises of estimating quantitatively, in the kinematical region of the present machines, the size of the contributions embodied in the BFKL equation (and, possibly, in sub-leading corrections to it).

The systematic use of high-energy factorization [5,6] has shown that these contributions can be recast in the form of an infinite resummation to all orders in $\alpha_S$ of classes of perturbative corrections to the anomalous dimensions and coefficient functions which enter in the renormalization group analysis. It is therefore possible to address phenomenological questions at small $x$ by incorporating enhanced higher-order corrections in the QCD evolution equations. On this basis, an attempt to implement QCD evolution at small $x$ beyond fixed perturbative order is in progress, and some results have already been published [7].

In this paper we recall the main theoretical inputs to this calculational programme, and discuss the impact of small-$x$ resummation on the predictions for the deep inelastic structure function $F_2$.

# 2. Scaling violation at small $x$

Let us start by recalling the implications of the BFKL analysis for the scaling violation in deep inelastic scattering. In QCD perturbation theory the amount of scaling violation is controlled by the anomalous dimensions $\gamma_{ab}$ ($a, b = q_i, \bar{q}_i, g$, $i = 1, \ldots, N_f$, $N_f$ being the number of active flavours) as functions of the QCD coupling $\alpha_S$ and the moment variable $\omega$ conjugate to $x$ in the Mellin-Fourier transform. It is known that in the limit $\omega \to 0$, which probes the small-$x$ region, the gluon anomalous dimensions $\gamma_{gb}$ behave like $\alpha_S/\omega$ at one-loop level. In higher loops, multiple gluon exchange actually gives rise to whole towers of poles, as follows

$$\gamma_{ab}(\omega, \alpha_S) = \sum_{k=1}^{\infty} \left(\frac{\alpha_S}{\omega}\right)^k A_{ab}^{(k)} + \sum_{k=0}^{\infty} \alpha_S \left(\frac{\alpha_S}{\omega}\right)^k B_{ab}^{(k)} + \mathcal{O}\left(\alpha_S^2 \left(\frac{\alpha_S}{\omega}\right)^k\right) \tag{1}$$

where the terms $A^{(k)}$ are referred to as the leading (L($x$)) series, the terms $B^{(k)}$ as the next-to-leading series (NL($x$)), and so on. The observation of Lipatov and collaborators [4] was that in fact, to the leading logarithmic accuracy L($x$), gluon radiation satisfies factorization properties in terms of real and virtual vertex operators in the colour space, which allow one to write an integral equation for the gluonic off-shell amplitudes, and ultimately to obtain the resummation of the L($x$) contributions to Eq. (1). The outcome is shown in Fig. 1, where different perturbative approximations to the leading anomalous dimension $\gamma_L$ are compared with the resummed result. The summation of the perturbative $\omega$-poles builds up a branch-point singularity at a value $\omega_L = 4 \ln 2 \, \alpha_S N_c / \pi$, $N_c$ being the number of colours (in Fig. 1 the moment variable has been rescaled in such a way that the branch point falls at 1/2). As $\omega$ decreases through $\omega_L$, the real part of $\gamma_L$ rises quickly until it saturates at the value 1/2. The study of the phenomenological consequences of this singularity structure has been the object of much effort in recent years, and is also the first question we would like to address in the next Section.

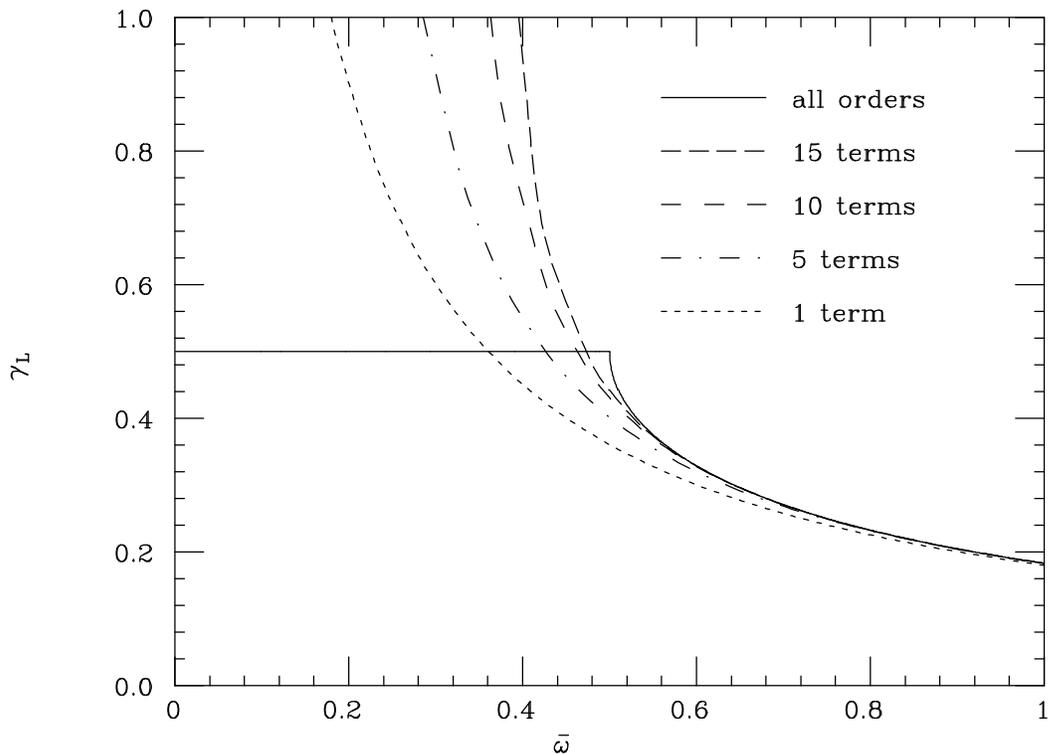

Figure 1: The BFKL anomalous dimension $\gamma_L$ (real part), as a function of the rescaled moment variable $\bar{\omega} \equiv \omega/(8 \ln 2 \, \alpha_S N_c/\pi)$.

So much for the leading-order behaviour. However, as in any leading logarithmic approximation, one needs to control corrections in sub-leading orders to estimate the accuracy of the theoretical predictions and carry out quantitative comparisons of the theory with experiment. The issue of the sub-leading corrections at small $x$ is particularly relevant if one wishes to describe the structure function $F_2$ at resummed level, because $F_2$ couples directly to quarks, and quarks start to contribute to the small-$x$ logarithmic expansion (1) only in next-to-leading order (NL($x$)). The calculation of the $\mathcal{O}\left(\alpha_S(\alpha_S/\omega)^k\right)$ contributions to the quark anomalous dimensions has been performed in Ref.[6], using a property of factorization of quark loops at fixed transverse momentum. The comparison of the resummed anomalous dimension and various perturbative truncations is shown in Fig. 2 for the case of the DIS factorization scheme. Note that next-to-leading resummation in the quark sector does not change the position of the singularity with respect to the leading-order case, but rather has the role of accelerating the approach to it, giving rise to positive definite corrections to the scaling violation. The second phenomenological question we want to address is therefore to what extent quark evolution affects the predictions for the structure functions at small $x$.

## 3. Numerical results for the structure function $F_2$

The discussion above can be summarized looking at the QCD evolution equations in

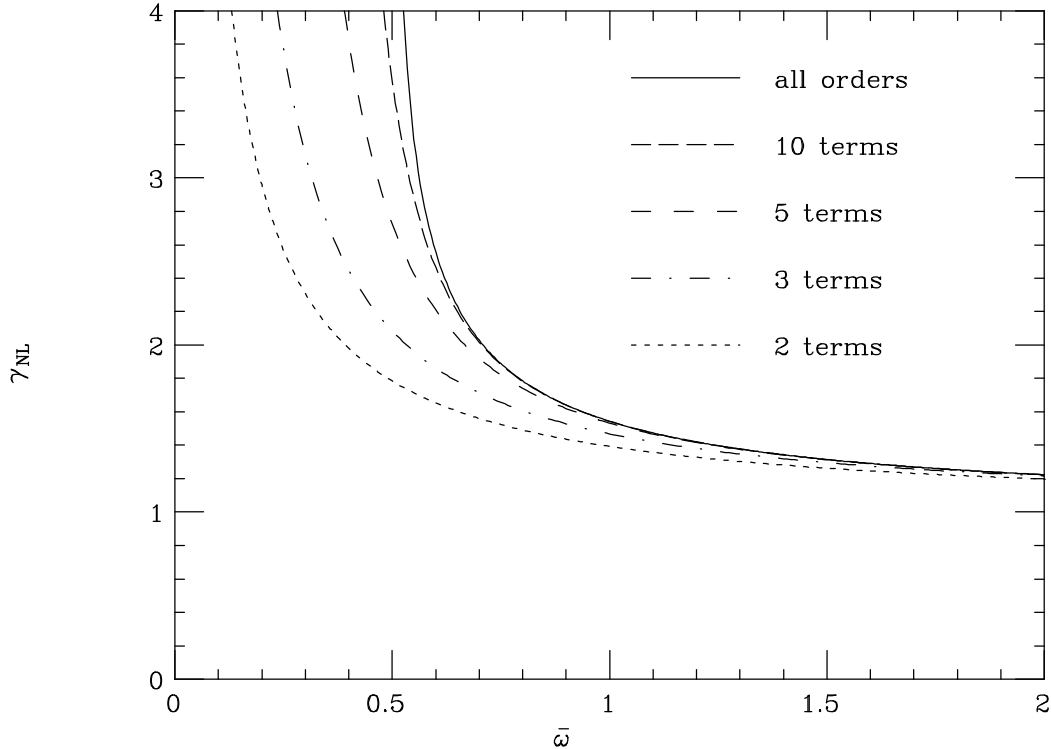

Figure 2: The quark anomalous dimension $\gamma_{NL}$ (real part) in the DIS factorization scheme, as a function of the rescaled moment variable $\bar{\omega} \equiv \omega/(8 \ln 2 \alpha_S N_c/\pi)$.

the flavour-singlet sector (flavour non-singlet components never get enhanced at $\omega \to 0$)

$$\begin{pmatrix} \dot{f}_S \\ \dot{f}_g \end{pmatrix} = \begin{pmatrix} \gamma_{SS} & \gamma_{Sg} \\ \gamma_{gS} & \gamma_{gg} \end{pmatrix} \begin{pmatrix} f_S \\ f_g \end{pmatrix} \quad , \tag{2}$$

where $f_g$ and $f_S$ are the gluon and the quark singlet density respectively. At small $x$, up to NL($x$) accuracy, the gluon entries of the matrix kernel are given by $\gamma_{gg} = \gamma_L + \gamma_\eta$, $\gamma_{gq} = (N_c^2 - 1)\gamma_L/(2N_c^2) + \gamma_\delta$, and the quark entries are $\gamma_{Sg} = \gamma_{NL}$, $\gamma_{SS} = (N_c^2 - 1)(\gamma_{NL} - \alpha_S N_f/(3\pi))/(2N_c^2)$. By $\gamma_\eta$, $\gamma_\delta$ we denote gluonic next-to-leading contributions, which are only known up to two loops at present.

The procedure we follow to study quantitatively the effect of enhanced higher-order contributions to QCD evolution at small $x$ consists of matching the resummed kernels $\gamma_L$, $\gamma_{NL}$ with fixed-order perturbation theory at one-loop [8] and two-loop [9] level, and solving the resulting evolution equations. Given a set of input distributions at a starting scale $Q_0^2$, this procedure allows one to determine predictions for the parton densities at another scale $Q^2$ which incorporate small-$x$ dynamics effects as well as renormalization group effects. In particular, the running of the QCD coupling is included in a straightforward manner up to two-loop level, much in the same way as in standard perturbative calculations. We stress that it is essential to set up a procedure in which both kinds of effects are taken into account consistently, especially to analyze the HERA region, where one may expect them to give contributions of comparable order of magnitude.

This matching procedure is obviously ambiguous by terms beyond two-loop order which are sub-leading at small $x$. In particular, this applies to corrections around $\omega = 1$, which control momentum conservation in higher loops. We choose to define the matching in such a way that the momentum sum rule is enforced to all loops. To this end, we consider two different prescriptions: one in which resummed higher-order expressions get multiplied by sub-leading factors of $(1 - \omega)$, and another one in which their values at $\omega = 1$ are subtracted off. One may interpret the band between the results obtained from these two different models as an estimate of the theoretical uncertainty on the resummed predictions due to unknown sub-dominant terms.

The results for the parton densities can then be combined with appropriate coefficient functions, also resummed to the corresponding level of accuracy [6], to obtain predictions for the deep inelastic structure functions. In Fig. 3 we show the results for the structure function $F_2$ [7] corresponding to the case of flat input parton distributions. We assume the set MRSD0' [10] as starting distributions at $Q_0^2 = 4$ GeV$^2$, and compare fixed-order and resummed evolution. The curves labelled L($x$) contain, in addition to the two-loop terms, the resummation of the BFKL contributions. For the curves containing also the next-to-leading resummation we use the label NLQ($x$), in order to remind the reader that only the quark sector is treated with next-to-leading accuracy, and therefore these predictions do not represent the full next-to-leading result. We show results corresponding to the two models for momentum conservation described earlier, the subtractive model being identified by the label $(H)$. For reference purposes we also report the 1993 ZEUS data [1].

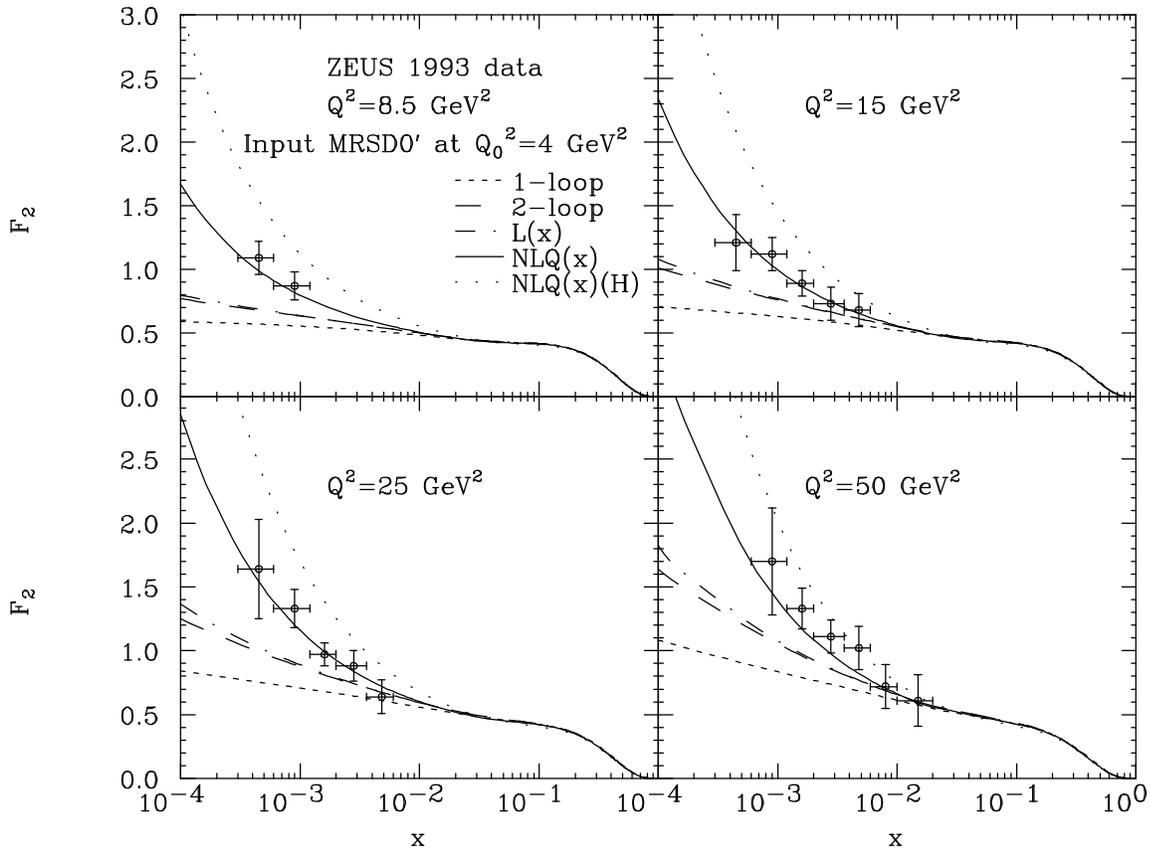

Figure 3: Resummed predictions for the structure function $F_2$.

The overall observation is that the impact of small-$x$ dynamics on the evolution of flat input distributions is large already in the HERA region. The resummed predictions (solid curves) are well above the two-loop ones (dashed curves). In particular, comparison of the $L(x)$ and $NLQ(x)$ curves shows that the largest contribution comes from the corrections to the quark anomalous dimensions. The structure function $F_2$ in the HERA region is only moderately affected by the BFKL terms, and on the other hand gets large contributions from quark evolution at small $x$.

Note that resummation effects set in over a relatively short evolution span in $Q^2$. Fig. 3 illustrates the fact that, starting with flat input distributions at $Q_0^2 = 4$ GeV$^2$, one can observe a growth in the structure function already around $Q^2 = 8.5$ GeV$^2$ as a consequence of perturbative resummation. The rise is just about as steep as the behaviour which is being observed in the HERA data. However, a word of caution is needed in interpreting these results. Observe that the resummed results are rather sensitive to the matching prescription. This can be regarded as an indication that the logarithmic expansion at small $x$ may still be subject to large corrections in the HERA region. One may then take a conservative attitude and interpret the difference between the two $NLQ(x)$ curves, obtained with the two extreme matching prescriptions, as an estimate of the theoretical uncertainty associated with the present resummed predictions.

As far as the $L(x)$ accuracy is concerned, it is worth recalling that results qualitatively similar to the ones presented here were obtained earlier in the literature using an approach based on the numerical solution of the BFKL equation supplemented with phenomenological models to include renormalization group effects (such as running coupling). For instance, one of the first studies on gluon evolution at small $x$ was performed by Kwiecinski in 1985 [11]. As to the effect of quark evolution, note that the onset of the behaviour discussed in this paper can already be seen numerically at fixed order in the study of Ref.[12]. In that case no resummation is involved, but one may note from the fixed-order predictions for $F_2$ that the $\mathcal{O}(\alpha_S^3/\omega^2)$ term in the quark anomalous dimension gives the dominant contribution to the rise at small $x$.

The presence of large higher-order corrections raises the question of the stability of the logarithmic expansion at small $x$. One way to estimate this is to compare the relative contributions to the structure function from different orders of approximation. To this end, we take the difference between the $L(x)$ result and the one-loop result, divided by the total, and then the difference between the $NLQ(x)$ result and the $L(x)$ result, again divided by the total. These ratios are plotted in Fig. 4 for a typical value of $Q^2$ ($Q^2 = 50$ GeV$^2$). We observe that in the HERA region both ratios stay below the value $1/2$, which corresponds, in each case, to the correction becoming as large as the lower-order result to which it is added, and are roughly of the same size. That is, resumming the next-to-leading logarithms at small $x$ does not give a much larger effect than that due to lower-order (two-loop terms, leading logarithms at small $x$) contributions. One may take these facts as indications that, despite corrections tending to be large in this regime, the series can still be controlled perturbatively. It is also worth mentioning that, as $Q^2$ increases, for any given value of $x$, the dashed curve takes over at high enough $Q^2$, that is, the normal perturbative behaviour is restored.

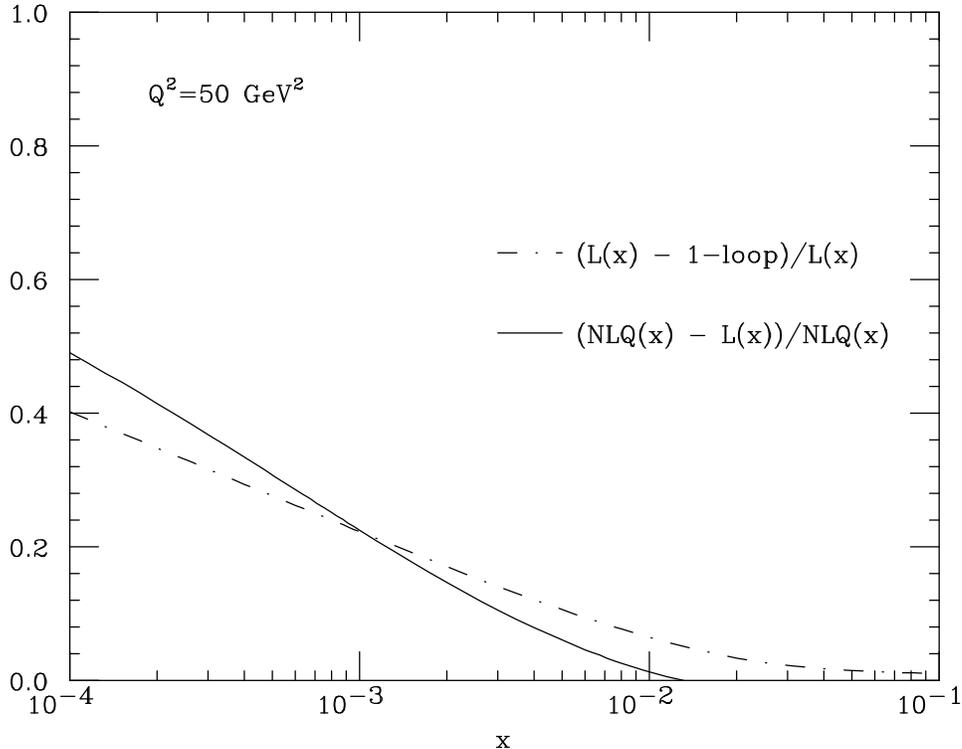

Figure 4: Relative contributions to the structure function $F_2$ at small $x$.

We have so far referred to the case of flat input distributions. The evolution of steep input distributions, on the other hand, is not so strongly affected by small-$x$ resummation of the kernels. We illustrate this in Fig. 5 for $Q^2 = 8.5$ GeV$^2$, where we assume the set MRSD$\_'$ [10] as starting distributions. We see that the difference between the resummed predictions and the fixed-order ones is less marked in this case. Observe also that, if the input is steeper, the sensitivity to different prescriptions for momentum conservation is reduced, and the associated uncertainty may be taken to be smaller. However, note that the 1993 HERA data [1,2] rather clearly disfavour an input as steep as MRSD$\_'$ (see Ref.[13] for an up-to-date review of parton distributions, and Ref.[14] for detailed fits, partially taking into account higher-order effects).

## 4. Conclusions

We have reported on a study of QCD evolution at small $x$ beyond two-loop order. Our approach is based on the high-energy factorization analysis, which allows one to recast the BFKL terms and corrections to them in the form of an infinite resummation of perturbative contributions to the QCD anomalous dimensions, and to match them with conventional renormalization group effects in a consistent way to all orders in the strong coupling $\alpha_S$.

The quantitative impact of small-$x$ resummation depends on the form of the input parton distributions. We find that there are sizeable effects in the case of flat input, and that, in particular, small-$x$ resummation can account for the rise of $F_2$ being observed at HERA. The effects are reduced if the input is steep.

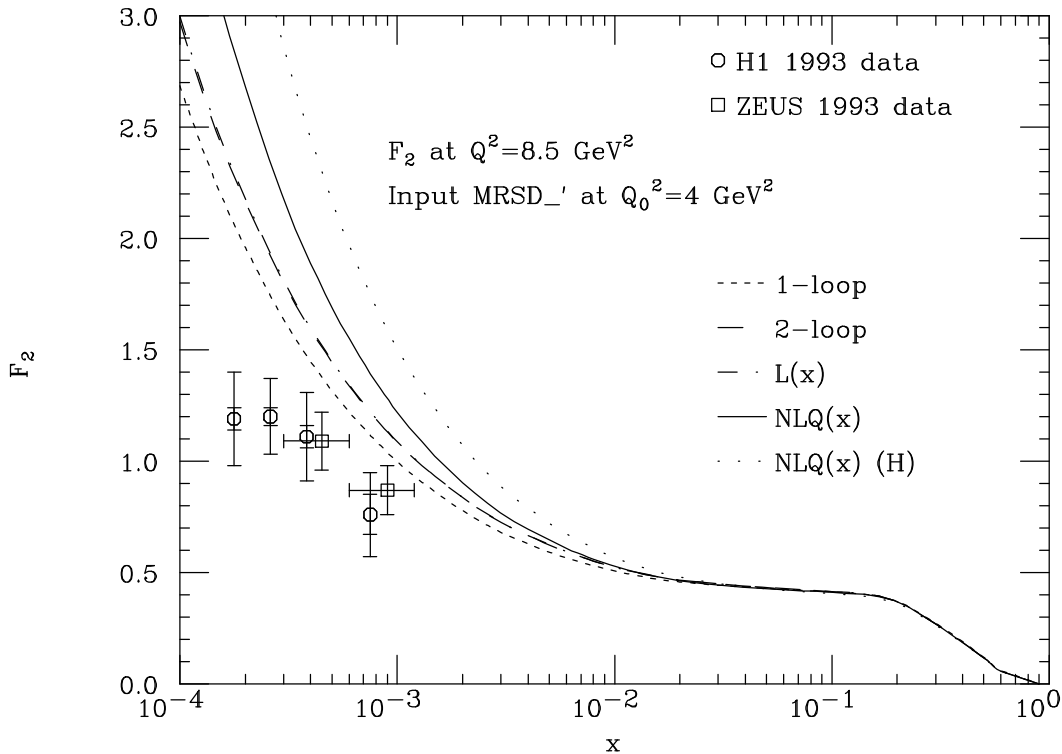

Figure 5: The case of steep input distributions. The different curves are as in Fig. 3.

The quality of the resummed predictions is at present affected by rather large uncertainties. On one hand, the calculation of the next-to-leading $\mathcal{O}\left(\alpha_S(\alpha_S/\omega)^k\right)$ corrections is not complete. On the other hand, models for momentum conservation in higher loops suggest that sub-dominant contributions may still be important in the HERA region.

**Acknowledgments.** The work reported in this paper has been carried out in collaboration with R. K. Ellis and B. R. Webber. This research is supported in part by the UK Particle Physics and Astronomy Research Council and the EC Programme "Human Capital and Mobility", Network "Physics at High Energy Colliders", contract CHRX-CT93-0357 (DG 12 COMA).